\documentclass[preprint,showpacs,amsfonts,amssymb,amsmath,12p]{revtex4}
\usepackage{times}
\usepackage{graphicx}

\begin{document}

\title{Casimir Friction I: Friction of a vacuum on a spinning dielectric}

\author{Yves Pomeau$^{1,2,3}$ (pomeau@math.arizona.edu) and David C. Roberts$^3$ (dcr@lanl.gov)}  

\affiliation{$^1$Department of Mathematics, University of Arizona, Tucson, AZ \\ 
$^2$ Laboratoire Physique statistique, Ecole normal Superierure, Paris, France \\
$^3$ Theoretical Division and Center for Nonlinear Studies, Los Alamos National Laboratory, Los Alamos, NM}

\maketitle

\section{Casimir Friction}
Quantum mechanics differs from classical mechanics in a number of ways. In particular there is a profound difference between a quantum mechanical ground state and a state of rest in classical physics. To illustrate, consider a physically empty cavity with boundaries that are made of a perfect conductor and thus reflect electromagnetic (EM) waves. Let us suppose that this system is at zero temperature, which is equivalent to saying that each mode of fluctuation of the EM field is in its quantum ground state. Because there are infinitely many modes, with a number density in energy space proportional to the three halves power of the modal frequency, the energy represented by these zero-point fluctuations diverges massively at high frequencies. This divergence arises from the part of the spectrum where the wavelengths are very short (in technical terms therefore, this is known as ``ultraviolet" divergence). The divergence should have a dramatic effect on the right-hand side of Einstein's equations of general relativity, which is proportional to this energy. And yet, even though the divergence of the energy is so dramatic, it is unknown if its effect on Einstein's equations describing large-scale gravitational effects has any measurable consequence at scales accessible to current experimental technology.  This is one of the greatest mysteries of present-day theoretical physics and little progress has been made since Pauli first noticed the incongruity.  

For physical phenomena where the absolute zero of energy can be arbitrarily assigned, however, this divergence is in principle irrelevant. Nevertheless it was realized by Casimir \cite{Casimir} (perhaps with prompting by Niels Bohr) that the difference between a quantum mechanical ground state and a classical state of rest could have experimentally verifiable consequences. Casimir reasoned that the shape of the cavity, such as that described above, defines its boundary conditions and, as a result, determines the cavity's energy content.  Combined with the fact that an energy differential would give rise to a force, he argued that a force should arise between a pair of parallel conducting plates held in an EM vacuum.  This Casimir phenomenon, where changing boundary conditions on zero-point fluctuations yields a force, is in principle limited neither to EM fluctuations nor to the measurement of forces between conducting plates, although these in combination are the most thoroughly investigated situation so far, both theoretically and experimentally. Below we review some new ideas in the field. 

We begin with the following observation: an object moving in an EM vacuum cannot feel any force if it moves at constant velocity; for a force would determine an absolute frame of reference and thus break the Lorentz invariance of the EM vacuum. When the object ``scatters" the zero-point fluctuations, no net force can be produced. However, it is possible to think of other situations where a net physical effect does arise from the scattering of zero-point fluctuations. One such situation is Casimir's parallel-plate configuration where the very presence of the second plate breaks the translational property of space for the first plate and a force is created. 

We examine another such situation below: a piece of dielectric rotating in an EM vacuum.  As has been known since the publication of Newton's Principia, rotation exists absolutely contrary to translation at constant speed. Therefore a force (in an extended sense) may be generated by the scattering of zero-point fluctuations off a dielectric rotating at constant angular speed.  The dielectric should be subject to a constant torque. 

Further, in a companion article \cite{enc}, we will explore a variant on Casimir's idea whereby a gaseous Bose-Einstein condensate at very low temperature takes the place of the EM vacuum.  (Our analysis should also be applicable to liquid helium at very low temperature.)  This alternative physical context is not bound by Lorentz invariance.  There is an absolute frame of reference for the condensate held fixed in space and, a priori, the possibility of a force generated by the motion of an object inside this condensate. While this situation has been the subject of investigation over the years, the discussion presented here takes a slightly different point of view. We show below that a force is generated by the scattering of zero-point fluctuations off an object moving at constant speed in a condensate. This force exists for any speed but it is of a special kind because it is reversible in some sense and is compatible with Landau's idea of a critical speed beyond which irreversible dissipation sets in. 

It should be noted that another type of ``Casimir" friction that is not discussed below occurs when a dielectric is moved at constant velocity relative to another dielectric.  Since the additional dielectric sets an absolute rest frame, a drag force exists but is extremely small (see \cite{yp2} for a discussion) and does not violate the Lorentz invariance of the EM vacuum.  For a review of this and other ``dynamic" Casimir effects, see \cite{kardar}.

\section{Casimir Friction I: Friction of a vacuum on a spinning dielectric}
\label{sec:frictionspinning}

This section reports on what has been presented in two recent papers, \cite{yp1} and \cite{yp2}, the first one introducing the general idea and the second one introducing a possible experimental set-up to check it.  We shall review the theoretical idea and refer the interested reader to the original reference for the experimental aspect. 

As said before, Newton very carefully considered the difference between translation at constant speed {\it{in vacuo}}, which cannot be detected in the co-moving frame, and rotation at constant speed which can, indicated by the curvature of the surface of a liquid at rest in the rotating frame, the well-known Newton's bucket. Consider a piece of matter rotating within a vacuum at rest.  Together, the piece of matter and the vacuum make a system that is not at equilibrium.  Naturally, there is a tendency for the system to relax to equilibrium. Indeed, as is well known, a rotating electric charge radiates classical EM waves, which carry away angular momentum. This is not a quantum phenomenon, although quantum effects change quantitatively the radiation of angular momentum. 

In contrast, the scattering of zero-point fluctuations by a rotating piece of dielectric is a purely quantum effect that generates a frictional torque on the dielectric. We sketch below the calculation of this torque. It relies on two ideas. The first involves the scattering of a plane wave by a rotating piece of dielectric, which we consider in the limit where this dielectric is much smaller than the wavelength of the incoming wave. Assuming then that this dielectric is an axissymmetric ellipsoid, one can find the induced dipole. Because this dipole rotates with the imposed angular frequency, the far field has frequency components at $\omega \pm \Omega$ and $\omega \pm 2 \Omega$, $\omega$ being the frequency of the incoming wave and $\Omega$ the angular velocity. The shift of frequency between the incoming and outgoing wave explains ultimately that some amount of angular momentum is radiated to infinity by the scattered wave. By conservation of angular momentum, this angular momentum should be balanced by a torque exerted on the rotating dipole. This torque is proportional to the square of the amplitude of the incoming wave. Therefore, once averaged over the quantum fluctuations, this mean-squared value is not zero and yields a net effect. 

Below we sketch the calculation by stressing its main steps (details can be found in \cite{yp2,yp1}).

Let ${\bf{P}}(t)$ be a time-dependent dipole. Assume it depends sinusoidally on time:

$$ {\bf{P}}(t) = \frac{1}{2} \left( {\hat {\bf{P}}} e^{i\omega t} +  {\hat {\bf{P}}}^* e^{-i\omega t} \right) \mathrm{,}$$ where * represents the complex conjugate; quantities with $\hat{}$ are complex (numbers or vectors) but time independent. 

Assuming the dipole is at position ${\bf{r}} = 0$, the electric field $\bf{E}$ generated by this dipole is given by an equation derived first by Hertz:

\begin{equation}
{\hat {\bf{E}}} = k^2 \frac{e^{ikr}}{r} \left[ {\hat {\bf{P}}}-  {\bf{n}} ( {\bf{n}}\cdot {\hat {\bf{P}}})\right] +   \left[ 3{\bf{n}} ( {\bf{n}}\cdot {\hat {\bf{P}}}) -  {\hat {\bf{P}}}\right] \left(\frac{1}{r^3} - \frac{ikr}{r^2}\right) e^{ikr}.
\label{eq:hertz}
\end{equation}
In this equation ${\bf{n}}= \frac{{\bf{r}}}{r}$ is the unit normal in the outward radial direction and $k$ is the wavenumber. 
The flux of angular momentum at infinity is derived from this expression  of the electric field by computing the Maxwell stress and its contribution to the flux of angular momentum. Integrating this flux across a large sphere surrounding the dipole, one finds the radiated torque $\bf {\Gamma}$ as a function of the polarization of the emitting dipole:

\begin{equation}
{\bf{\Gamma}} =  \frac{16 i \pi k^3}{3} {\hat {\bf{P}}} \times {\hat {\bf{P}}}.
\label{eq:torque}
\end{equation}
This polarization is itself a function of the amplitude of the incoming EM wave. In the limit where the physical size of the dipole is far less than the wavelength of this wave, this electric field can be considered as constant near the dipole. In that case there is a simple linear relation between the incoming field $ {\bf{E}}$ and the resulting polarization  $ {\bf{P}}$:
\begin{equation}
P_i =  A_{ij} E_j
\mathrm{.}
\label{eq:response}
\end{equation}
$A_{ij}$ is a rank-two tensor attached to the dipole, with the physical dimension of a volume. Indices $i$, $j$, etc. denote Cartesian components and summation over like indices is implied. 

For an ellipsoidal dielectric of uniform polarizability the tensor $A_{ij}$ is known explicitly. Suppose the ellipsoid is limited by a surface given by the Cartesian equation 
$$\frac{X^2}{a^2} +  \frac{Y^2}{b^2} + \frac{Z^2}{c^2} = 1 \mathrm{,}$$
$(X \mathrm{,}Y \mathrm{,}Z)$ being the system of coordinates attached to the ellipsoid and $a$, $b$ and $c$ being the constant parameters defining the surface. In this system the tensor $A_{ij}$ is diagonal and the$X$-component of the dipole induced by $E_X$ is
$$ P_X = A_{XX} E_X \mathrm{,} $$
where 
$$  A_{XX} = \frac{a b c }{3\left[ \frac{\epsilon}{\epsilon_1 - \epsilon} + m_{(X)}\right]}  \mathrm{.}$$ 
The quantity $m_{(X)}$ depends on $(a,b,c)$ and is given by the elliptic integral:
$$m_{(X)} = \frac{a b c}{2} \int_0^{+\infty}   \frac{\mathrm{d}\zeta}{(\zeta + a^2) R(\zeta)}  \mathrm{,}$$ 
where $R(\zeta) = \sqrt{(\zeta + a^2) (\zeta + b^2)(\zeta + c^2)}$. 
Things are made easier by considering a dipole with $Z$ as an axis of symmetry. Let ${\bf{N}}$ be then the unit vector along this axis of symmetry. The rank-two tensor $A_{ij}$ is the sum of the tensor $\alpha N_i N_j$ and an isotropic part $\beta \delta_{ij}$, where $\delta_{ij}$ is the Kronecker tensor. For this symmetric ellipsoid, $\alpha = A_{ZZ} - \frac{1}{2}(A_{XX} + A_{YY})$ and $\beta = A_{XX} = A_{YY}$, and the relation (\ref{eq:response}) between the incident electric field and the polarization becomes
\begin{equation}
{\bf{P}} = \alpha {\bf{N}}({\bf{N}}\cdot{\bf{E}}) + \beta {\bf{E}}
\mathrm{.}
\label{eq:response1}
\end{equation} 
The last term comes from the isotropic part of the polarization tensor and is not sensitive to changes of orientation induced by the rotation of the dipole. Therefore it does not induce any torque and will be discarded from now on. 

To estimate this torque, one assumes that the dipole rotates at angular speed $\Omega$ around an axis that make an angle of $\theta$ with the axis of symmetry of the ellipsoid. 
Let $z$ be the axis of rotation. By symmetry, the resistive torque has a non-zero component along $z$ only, which reads
\begin{equation}
\Gamma_z =  \frac{16 i \pi k^3}{3} \left(\hat {P}_x \hat {P}_y^* - \hat {P}_y \hat {P}_x^*\right)
\mathrm{.}
\label{eq:torquez}
\end{equation}
 This equation is valid for one frequency component of $ {\bf{P}}$ and the wavenumber $k$ entering in this expression is the one associated with it. Because of the time dependence of ${\bf{N}}$ due to the rotation at angular speed $\Omega$, for an incoming field ${\bf{E}}$ at frequency $\omega$ the polarization gets frequencies $\omega \pm \Omega$ and $\omega \pm 2 \Omega$. This combination of frequencies results in a non-zero torque $\Gamma_z$, as derived from (\ref{eq:torquez}) :
\begin{equation}
\Gamma_z = \alpha^2 \frac{E_x^2}{6c^3} \sin^4(\theta)\left[(\omega - 2\Omega)^3 - (\omega + 2\Omega)^3 \right] + 
 \alpha^2 \frac{2 E_z^2}{6c^3} \sin^2(\theta)\cos^2(\theta) \left[(\omega - \Omega)^3 - (\omega + \Omega)^3 \right] 
\mathrm{.}
\label{eq:torquezexp}
\end{equation} 
This equation shows that without rotation (i.e. $\Omega =0$) there is no torque. In the realistic limit $\Omega<<\omega$, equation (\ref{eq:torquezexp}) becomes:
\begin{equation}
\Gamma_z = - \omega^2 \Omega\left[\alpha^2 \frac{2 E_x^2}{c^3} \sin^4(\theta) +  
4  \alpha^2 \frac{ E_z^2}{c^3} \sin^2(\theta)\cos^2(\theta) \right]
\mathrm{.}
\label{eq:torquezexplim}
\end{equation}
The coefficient of polarization $\alpha$ has been taken as independent of the frequency. However, such a dependence could be taken into account. The torque given by equation (\ref{eq:torquezexplim}) is proportional to the square of the field of the incoming wave. Because the mean value of a quantity proportional to the square of a quantum fluctuation is not zero in general, as the regular Casimir effect shows, one may expect a torque due to the scattering of zero-point fluctuations. The order of magnitude of this torque is found by adding the contributions to $\Gamma_z$ from all modes of fluctuations of the QED vacuum, each mode having its own ground state amplitude. The contribution of a mode of frequency $\omega$ to $|E|^2$ is of order $\frac{\hbar \omega}{2 V}$, where $V$ is the volume of the system and assumed to be very large. In the frequency interval $[\omega,  \omega + {\mathrm{d}}\omega]$ the number of modes of the EM field is $\frac{ 8\pi V \omega^2 {\mathrm{d}}\omega}{c^3}$. In the limit where $\Omega$ is small, the sum of all contributions to the Casimir torque $\Gamma_C$ has a magnitude on the order of $\Omega \int_0^{\infty} {\mathrm{d}}\omega \alpha^2 \frac{\hbar \omega^5}{2c^3}$. As is often the case in this field, this integral diverges massively at large frequencies. There is a natural cut-off because of the assumption that the wavelength of the incoming wave is much bigger than the size of the dipole. Once this cut-off is put into the integral over the fluctuation modes, one finds for  $\Gamma_C$ the estimate
\begin{equation}
\Gamma_C \sim \hbar \Omega \left(\frac{\alpha}{a b c}\right)^2.
\end{equation}
The quantity $\frac{\alpha}{a b c}$ is dimensionless and of order $1$, although $\hbar \Omega$ has the dimension of a torque and also of an energy. Assuming that the resistive torque on each molecule of a rotating solid made of anisotropic molecules add together, the reference \cite{yp2} presents a possible experimental setup where this torque could be measured.  The idea underlying this experiment is to transfer the torque to a non-rotating piece of dielectric where it would become a static torque, much easier to measure than the torque on a rotating piece of matter.

\vspace{1 in}
{\large Mathematical Subject Classifications}
\begin{itemize}
\item 81V10 Electromagnetic interaction; quantum electrodynamics
\item 81U99 Scattering Theory
\item 82C10 Quantum dynamics and nonequilibrium statistical mechanics (general)
\end{itemize}

\end{document}